%% file: harbeck.tex
\documentclass{aa}

\usepackage{graphicx}
\usepackage{txfonts}
\usepackage{natbib}
\usepackage{xspace}

\newcommand{\ngc}{NGC\,7006\xspace}
\newcommand{\scnb}{S(3839)\xspace}
\newcommand{\scnr}{S(4241)\xspace}
\newcommand{\sch} {CH(4300)\xspace}

\begin{document}

\title{CN Variations in NGC 7006}
                                                                                
   \subtitle{}
                                                                                
   \author{Daniel Harbeck
          \inst{1}
          \and
          Graeme H. Smith\inst{2}
          \and
          Eva K. Grebel\inst{1}
          }
                                                                                
   \offprints{D. Harbeck, \email{dharbeck@mpia.de}}
                                                                                
   \institute{Max-Planck-Institut f\"ur Astronomie,  K\"onigsstuhl 17,
               69117 Heidelberg, Germany
            \and
               OCO/Lick Observatory, University of California, CA 95064, USA
             }
                                                                                
\date{Received <date> / Accepted <date>}
                                                                                
\abstract{
Rotationally induced mixing with subsequent dredge-up of
nucleosynthesized material is discussed as a second parameter of the
horizontal branch morphology in globular clusters. CNO abundances have
been proposed as tracers of the dredge up of processed material.
\ngc is a prominent example of a second parameter GC: Its HB
morphology is too red for its metallicity. We present spectroscopic
measurements of CN molecular band strengths S(3839) and CH band
CH(4300) strengths for 12 giants in \ngc to test rotationally-driven
mixing as a second parameter in this cluster. Our observations reveal
(i) a scatter in star-to-star CN absorption strengths with the same
amplitude as seen in other GCs of the same metallicity, but different
HB morphologies; (ii) a possible continuous distribution of CN
absorption strength with a preference for CN-enriched stars, and (iii)
a possible weak radial gradient in the number ratio of CN-strong and
CN-weak stars.  We argue against the hypothesis that CN-variations are
directly correlated with the second parameter effect of the HB
morphology. However, the small sample of stars measured in \ngc
prevents us from drawing firm conclusions. Finally, we identify one
star of our sample as a foreground dwarf carbon star.

 }
                                                                                
\maketitle

\section{Introduction}

The great ages ($> 10$~Gyr) of the globular clusters (GCs) in the
Milky Way (MW) ensure that only low-mass ($M\le0.9M_\odot$) stars are
present on the main-sequence, while the post main-sequence stars have
roughly the same initial mass for a given metallicity. According to
the Vogt-Russell theorem, the initial conditions of a star define its
subsequent evolution. Among MW GCs of the same metallicity therefore
all post main-sequence stars might be expected to evolve in the same
way, provided that they are coeval.

Such expectations are not in accord with observations: (i) While stars
of a given GC are homogeneous in the [Fe/H] abundance, differences in
light elements such as C, N, O, Na, and Al among GC red giant branch
(RGB) stars are a hint of additional processes governing their
evolution that are not included in standard stellar models.
Alternatively, the assumption of the same initial conditions on the
main sequence may be invalid. (ii) MW globular clusters of the same
metallicity show very different morphologies of the horizontal branch
(the ``second parameter effect''). Additional parameters are required
for a description of post main-sequence stellar evolution. In this
paper we investigate through a study of \ngc\ whether the second
parameter determining the HB morphology might be related to the
abundance inhomogeneities among the RGB stars.

\subsection{Morphology of the horizontal branch}

The helium core-burning sequence of a GC, i.e., the horizontal branch
(HB), is populated by stars of the same initial mass, and typical
present-day masses of $M\sim0.6\,M_\odot$. The HB morphology can be
quantified as follows \citep{lee94}:
\begin{eqnarray}
HB = \frac{\#B-\#R}{\#B+\#V+\#R},
\end{eqnarray}
where \#B and \#R are the number of HB stars bluer or redder than the
instability strip, respectively; \#V is the number of variable stars
in the instability strip.  Predominantly due to differential mass-loss
in the RGB phase prior to ignition of helium core-burning, an extended
region will be populated along the HB with stars of different envelope
masses. Among the MW GCs, there is a general dependence of the HB
morphology on cluster metallicity: On average, metal-poor GCs tend to
produce HB stars with high effective temperatures (blue HB stars),
while metal-rich GCs tend to form red HB stars with lower effective
temperatures.

However, it is known that other parameters must affect the evolution
of stars to the HB (e.g., \citealt{sandage60}), since there are GCs
that do not strictly obey this metallicity-HB-morphology relation. For
example, some GCs show excessively red HB morphologies at a fixed low
metallicity (e.g., \ngc, \citealt{sandage67}). Despite numerous
studies, the riddle of the second parameter in globular clusters still
lacks a comprehensive solution. For instance, age, helium abundance,
CNO abundances, or stellar rotation and subsequently induced mixing
might affect the HB morphology (as systematically investigated by,
e.g., \citealt{lee94}). In general, these parameters regulate the
evolution of a main-sequence star along the RGB and onto the HB by
changing the mass of the stellar envelope, the mass of the helium
burning core, the opacity of the stellar material, or a mixture of
these parameters.

Following the arguments of \citet{lee94}, we briefly summarize how
different stellar parameters can change the evolution of stars to the
horizontal branch. In this paper we test one particular scenario in
which internal abundance spreads in light elements in GCs could be
linked to the problem of HB morphology. We concentrate on the outer
halo cluster \ngc as it shows a prominent second parameter effect: At
a metallicity of [Fe/H]$=-1.6$~dex its HB morphology is redder than
that of other GCs of similar [Fe/H], such as M\,3, M\,13, or M\,10
(see Tab.~\ref{tab_7006_cn}).  We next discuss how the effects of
different parameters on the HB morphology have been explored for \ngc.

\begin{itemize}
  
\item {\bf Age:} Age is a promising candidate to be {\em the} second
  parameter effect: younger GCs would produce redder HBs than older
  GCs of the same metallicity. For instance, age differences measured
  with differential age-dating techniques from the main-sequence
  turn-off (MSTO) might be able to reproduce the differences in the HB
  morphologies of the pair NGC\,288/NGC\,362 at a metallicity of
  [Fe/H]$=-1.2$~dex, where NGC\,288 would be $2\pm1$~Gyr older than
  NGC\,362 \citep{catelan01} or for the pair M\,3/M\,13 at a
  metallicity of [Fe/H]$=-1.6$~dex, where M\,13 would be
  $1.7\pm0.7$~Gyr older \citep{rey01}.  \citet{buonanno91} found a
  ``normal'' age for \ngc from the difference in luminosity of the HB
  and the main sequence turn-over.  \citet{vdbergh98} and
    \citet{stetson99} found a general trend of outer halo GCs being
    one to two Gyr younger than the inner halo GCs. However,
    systematic differences in the $\alpha$-elements might also produce
    such a distance-age relation. If this general trend applied to
    \ngc, age might be indeed a viable second parameter of the HB
    morphology.  Unfortunately, due to its large distance, no
  precise relative age-dating of \ngc this cluster is available in the
  literature.

\item {\bf CNO abundance: }
  CNO abundances affect both the energy production rate on the
  main-sequence as well as the stellar opacity; as a net-effect,
  increased CNO abundances lead to redder horizontal branches.
  Especially in the case of \ngc the total CNO element abundance has
  been discussed as a potential second parameter: \citet{mcclure81}
  found very strong G band CH absorption in the RGB stars of this
  cluster, which might reflect a higher total abundance of CNO
  elements. Nevertheless, \citet{cohen82} could not confirm an
  overabundance of CNO elements in \ngc. These authors found that the
  infrared CO absorption band strengths of RGB stars are intermediate
  between those of M\,3 and M\,13; both GCs have bluer HB morphologies
  than \ngc. Finally, \citet{wachter98} found a depletion of carbon in
  \ngc giants. This is consistent with the existence of internal
  C$\rightarrow$N processing where nitrogen is nucleosynthesized by
  the CNO cycle in the stellar interior at the cost of carbon and
  oxygen. Internal mixing processes could bring this material up to
  the stellar surface. These authors conclude that if the reduced
  carbon abundance reflects a reduced total CNO element abundance,
  then the total amount of CNO is not likely to be responsible for the
  second parameter effect in \ngc. However, deep mixing can
  significantly deplete the surface carbon abundance of RGB stars
  while preserving the total C+N+O.Thus, carbon alone might be an
  improper proxy for the total CNO abundance of \ngc. If CNO
  enhancement in \ngc would be a second parameter of the HB
  morphology, increased CN and/or CH strength compared to GCs of bluer
  HB morphology but the same [Fe/H] would be expected.
  
\item {\bf Rotational mixing as a second parameter: } Understanding
  the effect of rotation on stellar evolution is of particular
  difficulty, since transport of angular momentum in the interior of
  stars adds extra challenge. Rotation can affect the later stellar
  evolution in several ways: rotation would delay the core helium
  flash, thus leading to higher core masses, resulting in hotter HB
  stars. Additional mass-loss due to reduced equatorial surface
  gravity in the later RGB phase would result in lower-mass stellar
  envelopes, also producing bluer HB stars. \citet{norris81a}
  suggested a scenario in which stars with sufficiently high angular
  momentum would be subject to rotationally-induced interior mixing
  (e.g, the detailed investigation by \citealt{sweigart97}). Such
  stars with increased internal mixing could dredge-up CNO-cycled, or
  incompletely CN-cycled or ON-cycled, material, altering the chemical
  composition of their atmospheres. At the same time, the increased
  angular momentum could lead to increased core masses as well as
  increased mass-loss during the RGB evolutionary phase, thus leading
  to blue HB stars. Slow rotators on the other hand --- appearing
  CN-poor --- would have lower mass loss and would therefore end up as
  red HB stars. {\em In this scenario a correlation between CNO
    surface abundances of RGB stars in a GC and the HB morphology
    would be expected.} In the case of M4 \citet{norris81a} found a
  similarity between the bimodal distribution functions of color along
  the HB and the CN absorption strengths of RGB stars, which would be
  consistent with this picture. The scenario in which the angular
  momentum distribution of stars in a GC would form a second or third
  parameter of HB morphology seemed promising.  However, recent
  measurements of stellar rotation speeds in M\,13 and M\,15
  \citep{behr00a,behr00b} identified the red HB stars as having higher
  mean rotational speeds (as found from the distribution of $v\sin i$)
  than the blue HB stars. In the rotational mixing scenario the
  opposite distribution would be expected. These observations could
  still fit a rotational second parameter scenario if increased
  mass-loss due to increased rotation speed would also imply increased
  loss of angular momentum, e.g., via stellar winds.
\end{itemize}

Rotational mixing as a second parameter leads one to expect a relation
between horizontal branch morphology and the distributions of the CNO
element abundances among GC red giants. We use spectroscopic
measurements of CN and CH bands of RGB stars as a tool to trace the
dredge-up of material that has been nucleosynthesized through the
CNO-cycle of hydrogen burning. As nitrogen is the minor component in
stellar atmospheres compared to C, an enrichment of N at the cost of C
by the CNO-cycle will result in enhanced molecular CN formation.

We aim to test the rotational mixing hypothesis by comparing both the
range of CN strengths exhibited by RGB stars, and the relative
fraction of CN-rich and CN-poor stars, in \ngc and a set of other GCs
having metallicities of [Fe/H]$ \sim -1.6$~dex but different HB
morphologies. For some of these GCs, measurements of CN absorption
strength are available in the literature. In the scenario of deep
mixing one would expect, with metallicity as a fixed parameter, a
correlation between the HB morphology of a globular cluster and the
star-to-star CN abundance variations within a cluster.  Besides the
strong ``global'' second parameter at work in \ngc,
\citet{buonanno91} found evidence for an ``internal'' second parameter
effect within it: The morphology of the HB becomes bluer with
decreasing radial distance. Unless one accepts the unlikely scenario
of an age-spread of order of a few billion years within a globular
cluster, such an internal second parameter effect provides evidence of
additional {\em third} parameters other than age, such as
environment. It will be interesting to test if this internal HB
gradient is also reflected by a gradient in the CN band strengths. We
note that such ``internal'' 2nd parameter effects have also been
observed in the old field populations of dwarf spheroidal galaxies
\citep{hurley99, harbeck01} and appear there correlated with the RGB
color.

\subsection{New CN and CH Spectroscopy of NGC 7006}

Although \ngc is an extreme example of the second-parameter effect in
GCs, it is poorly studied. Only a few measurements of CN absorption
strength are available in the literature. The major difficulty of this
object is its large distance (57.4~kpc), and only the tip of the RGB
has been explored spectroscopically so far. For example, from
high-resolution Keck spectra of six stars in \ngc \citet{kraft98}
detected a Mg-Al anti-correlation, which is a strong indication of
material nucleosynthesized by hydrogen-shell burning. A common tool
for investigating star-to-star inhomogeneities in the CNO elements is
spectroscopy of the 3883 \AA\ CN and 4300 \AA\ CH molecular absorption
bands, which does not require the high resolution needed for atomic
absorption line studies. The reduced telescope time required has to be
traded off against the loss of detailed information on individual
elements.  In this paper we report measurements of an index denoted
S(3839) for a sample of red giants in \ngc. This index has been used
in a number of CN studies. We compare the results with those for other
GCs of similar metallicity.

\section{Data and Reduction}

During several observing campaigns in the years 1996, 1997, and 2000,
spectra of 12 stars among the bright end of the RGB of \ngc were
observed with the KAST spectrograph on the Lick 3m Shane telescope.
The grism No.~1 was used, resulting in a wavelength coverage from
$3200$\,\AA\ to $6500$\,\AA\ at a dispersion of $2.8$\,\AA/pixel. A
$1200 \times 400$ Reticon CCD with a pixel size of $27\,\mu \rm{m}$
and $6\,\rm{e}^-$ readout noise was used as detector.  Per observing
run typically 20 bias observations and 20 dome flat fields were
taken. HeHgCd arc lamp exposures for wavelength calibration and flux
standard star spectra were obtained. A logbook of the observations is
given in Table~\ref{tab_7006_obslog}; the identification of the
observed stars is the same as in \citet{sandage67}.

The data were reduced separately for each observing campaign using
IRAF\footnote{IRAF is distributed by the National Optical Astronomy
Observatories, which are operated by the Association of Universities
for Research in Astronomy, Inc., under cooperative agreement with the
National Science Foundation.} data reduction routines. The KAST CCD
readout system automatically subtracts a bias level from the images
estimated from the overscan line. We found that the illumination level
on the chip affects the estimate for the overscan level (in particular
for highly illuminated flat fields). This effect leads to an
overestimation of the overscan level in some images. Since the
subtracted overscan is stored in the last column of each KAST CCD
image, we added back the overscan level. All bias frames were combined
and this masterbias was subtracted from all frames. In a subsequent
step the flat fields were combined. A polynomial of $25$th order in
the dispersion direction was used to eliminate both spectral
sensitivity and the energy distribution of the quartz lamp from the
flat field. The normalized flat field was used to flatten all data.

The long exposure spectra of \ngc are affected by cosmic ray hits. We
found that the best way to remove them was to use the {\tt
lacos\_spec} task \citep{dokkum01}. Most of the stars were observed
several times, and we coadded the available CCD frames to a final
deep, almost cosmic-ray-free image. The spectra were extracted with
the {\tt specred} package in IRAF. From the typically five flux
standard stars observed per campaign the spectral sensitivity function
was determined and used to flux-calibrate the stellar spectra. Note
that the spectra are only calibrated with respect to relative spectral
sensitivity, not to absolute flux.

\begin{figure}
 \resizebox{\hsize}{!}{\includegraphics{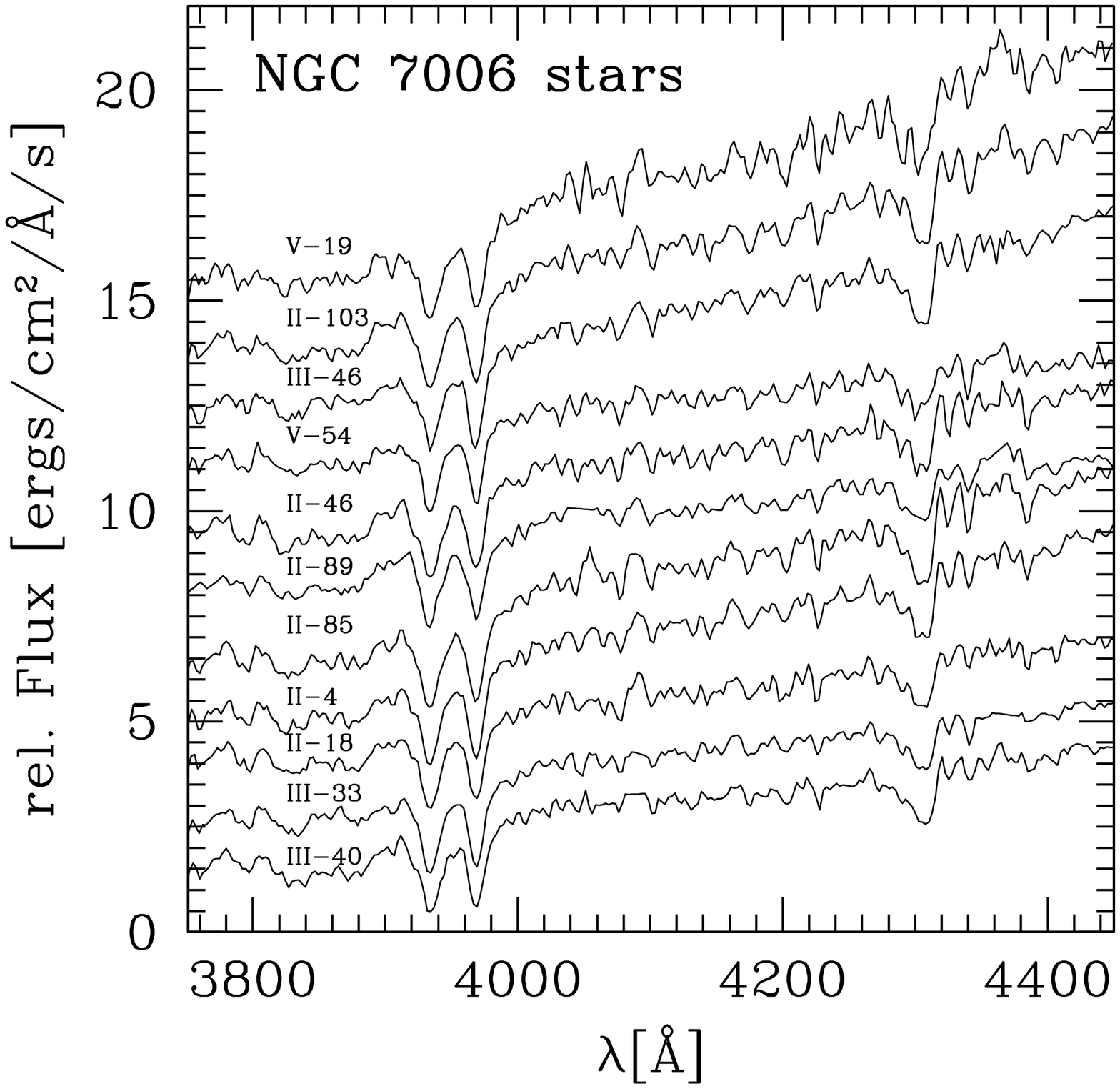}}
 \caption{Spectra of all stars in our sample; the peculiar star III-1
   is excluded. The spectra are normalized by their flux at 4200\AA,
   and they are shifted vertically for clarity.}
 \label{fig_7006_allspectra}
\end{figure}

For each spectrum we measured the centers of the Ca II H+K lines with
the {\tt rvidlines} task in the {\tt rv} package under IRAF to
determine the radial velocities of the stars.  Since we want to
measure spectral indices in the rest frame wavelength, we corrected
the spectra for the measured Doppler shift. The final extracted
spectra of all stars are plotted in Fig.~\ref{fig_7006_allspectra}.

\begin{figure}
 \resizebox{\hsize}{!}{\includegraphics{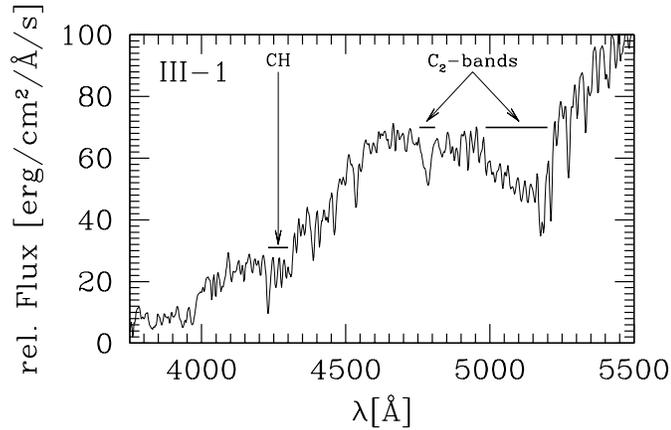}}
 \caption{Spectrum of the peculiar star III-1, which is probably a
 foreground dwarf carbon star. Note the strong absorption bands due to
 CH and C$_2$.}

 \label{fig_7006_spectrumiii1}
\end{figure}

Additionally we measured relative radial velocities via
cross-correlation against the star II-4 using the {\tt fxcor} task in
the IRAF {\tt rv} package. We found a nearly Gaussian distribution of
relative velocities with a dispersion of $\sim50$~km/s,  which
  reflects the uncertainty in the determination of the radial velocity
  at the given spectral resolution rather than the internal dispersion
  of \ngc.  But there is one significant outlier: The star III-1 was
found to have a relative radial velocity of $+245$~km/s compared to
II-4. We conclude that all stars in our sample but III-1 are members
of \ngc.  The spectrum of III-1 is peculiar and dominated by prominent
molecular bands (see Fig.~\ref{fig_7006_spectrumiii1}).  According to
its features (strong C$_2$ and CH absorption bands and strong Mgb
lines) we identify this star as a foreground dwarf carbon star.

\subsection{Spectral Indices}

From the spectral energy distribution we measured the strength of
absorption by the CN molecule at the $3883$\,\AA\ band. The CH
absorption strength was measured from the $4300$\,\AA\ G-band.  The
spectral indices \scnb and \sch measure the flux depression in the
molecular bands relative to a nearby comparison passband. The indices
are defined according to \citet{norris81}, but we modified the
definition of the CH index comparison passband to avoid strong sodium
sky emission lines. The CN index definition remains unchanged,
allowing a direct comparison to results of other studies:

\begin{eqnarray}
\rm{S}(3839) & = & -2.5 \log \left(\frac{\sum_{3846}^{3883}F_\lambda}
{\sum_{3883}^{3916}F_\lambda}\right)\\ \rm{CH}(4300) & = & -2.5 \log
\left(\frac{\sum^{4320}_{4280}F_\lambda}
{\sum_{4250}^{4280}F_\lambda+\sum_{4320}^{4340}F_\lambda}\right)
\end{eqnarray}

Larger index values correspond to increased molecular band absorption.
An additional spectral index \scnr to measure CN absorption in the
molecular band at $4215$\AA\ is not considered here, since the
absorption in this band is very low for metal-poor stars. Most of the
stars were observed several times. For two representative stars we
tested how well the individual measurements of the \scnb index would
reproduce the measurement on the combined spectrum. The typical
scatter in \scnb among the short exposures is of order
$\sigma$\scnb$=\pm 0.03$\,mag. To account for different exposure times
of the stars, we determine the uncertainty of the CN and CH indices
based on the Poisson noise of the number counts within the index
passbands,  resulting in error estimates that are of similar
  magnitude in both indices. The measured indices as well as
photometry from \citet{sandage67} are listed in
Table~\ref{tab_7006_indices}. In the following we will transform the
apparent magnitudes of \ngc stars to absolute magnitudes using a
distance modulus of (m-M)$_V=18.24$ and correcting for a reddening of
E(B-V)=0.05 \citep{harris96}

\section{Results} 

\begin{figure}

\resizebox{\hsize}{!}{\includegraphics{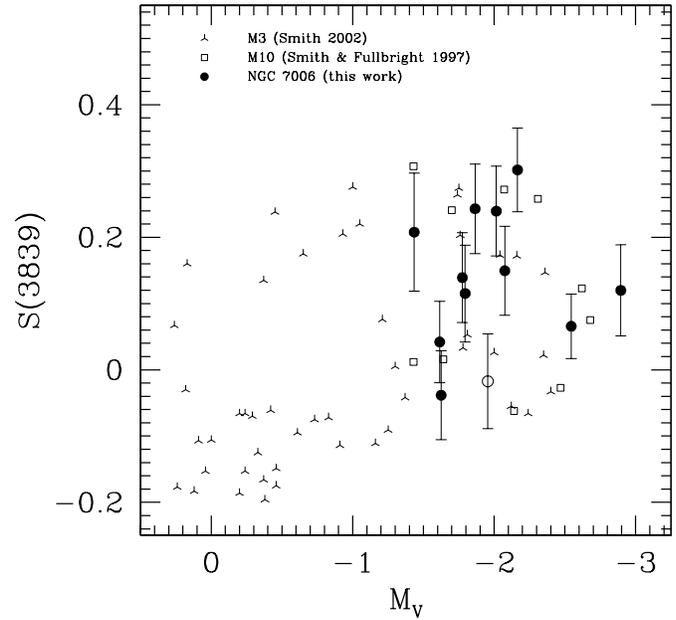}}

\caption{Comparison 
 of \scnb absorption strength in \ngc\ (filled circles), M\,3 (stars)
 and M10 (open boxes). In the magnitude range covered by our
 observations the scatter in the \ngc CN absorption strength compares
 well to those in M\,3 and M\,10. There is no clear bimodal signature
 in \ngc as is visible in M\,3. The foreground dC star III-1 is
 plotted with an open circle.}
\label{fig_7006_m3plot}
\end{figure}

The formation of CN and CH molecules is affected by temperature as
well as by gravity effects. Since our sample spans only a small range
in color (temperature) and luminosity, this dependence of the molecule
formation cannot easily be determined. We compare the \ngc \scnb
strengths to those of the well studied GCs M\,3 and M\,10, which have
the same metallicity as \ngc. \citet{smith02} homogenized CN \scnb
measurements of stars in M\,3 from numerous sources in the
literature. For M\,10, we use the measurements of \citet{smith97}. One
result of the \citet{smith02} study is a good description of the
dependence of the CN absorption strength on the luminosity/temperature
for stars on the upper RGB. A prominent feature is the decrease of CN
absorption strengths for stars brighter than $M_V\le-1.8$~mag. In
Fig.~\ref{fig_7006_m3plot} we compare our measurements of the
$3883$\,\AA\ CN band for \ngc (filled circles) with the literature
data for M\,3 (tripods) and M\,10 (open squares). The temperature
effect on the CN formation efficiency is clearly visible in the M\,3
sample; with increasing luminosity (=decreasing surface temperature)
the formation of CN becomes more efficient. At the turnover at
$M_V=-1.8$~mag the CN formation is reduced for increasing luminosities
as described by \citet{smith02}. We observed only the brightest stars
in \ngc, in particular all stars are at the luminosity of or brighter
than the CN-turnover luminosity. Two bright (and red) stars in our
sample are identified variable stars; at their location in the
color-magnitude diagram the risk of AGB star contamination is high,
and for a comparison between M\,3, M\,10, and \ngc we favor
concentrating on the stars closer to the CN-index turn-over region.

\subsection{Range of CN absorption strengths}

The range of CN absorption strengths at a given magnitude in \ngc (see
Fig. \ref{fig_7006_m3plot}) is comparable, within the errorbars, to
the range covered by the globular clusters M\,3 and M\,10. Both
comparison clusters have metallicities comparable to \ngc (see
Tab.\ref{tab_7006_cn}), but bluer HB morphologies (HB-values of
0.08 and 0.98, respectively).  There is no clear indication in
\ngc of a bimodal distribution of CN absorption strengths as
is observed in M\,3, M\,10, and in other globular clusters.  A
continuous distribution of S(3839) absorption strengths is possible
for \ngc, although the small size of our star sample prohibits a
unique determination of the CN distribution function.

In the scenario of rotational mixing as the second parameter we would
expect stars in GCs with blue HB morphologies to have, on average,
higher angular momentum than stars in GCs with red HB morphologies. We
consequently expect that stars in \ngc should be slower rotators than
those in M\,3 or M\,10. Under this assumption, stars in \ngc would
experience less rotationally induced mixing, and less
nucleosynthesized material should become dredged-up to the stellar
envelope. If rotation were primarily responsible for determining both
HB morphology and CN inhomogeneities, a smaller range in CN absorption
strength and/or a preference for stars with low CN abundances would be
expected in \ngc. Our observation of a range of S(3839) values in
\ngc that is comparable to those seen in GCs with bluer
HB morphology contradicts --- within the limitations of the small
sample --- the predictions of this scenario.

\subsection{Ratio of CN-strong/CN-weak stars}

We cannot identify a correlation between the range of CN absorption
strengths in \ngc, M\,3, and M\,10 and the HB morphology.  In a second
attempt we investigate the ratio by number
$r=\frac{\#CN-strong}{\#CN-weak}$ of CN-strong to CN-weak stars in GCs
with metallicities comparable to \ngc: In the scenario of rotationally
induced mixing the distribution of CN absorption strengths should be
weighted to CN-poor stars in GCs with low mean stellar angular
momentum (and relatively red HB morphologies), while CN-strong stars
should be preferred in GCs with high mean stellar angular momentum
(and bluer HB morphologies).

For GCs with bimodal distributions of CN absorption strengths a clear
classification into CN-strong and CN-weak stars is possible. We
collected values of $r$ for GCs with metallicities comparable to
\ngc where data are available from the literature
in Table~\ref{tab_7006_cn}. The metallicities and HB morphologies are
taken from \citet{harris96}.  In \ngc the classification of CN
absorption strengths is ambiguous: if we consider all
\ngc stars with magnitudes $M_V\ge-2.4$ and $S(3839)\ge0.1$ as
CN-strong (see Fig.\ref{fig_7006_m3plot}), we obtain a number ratio of
$r=\frac{7}{2}=3.5$. Ignoring stars with intermediate S(3839)
absorption strengths ($0.1\le S(3839)\le 0.18$), results in
$r=\frac{4}{2}=2$.

\begin{figure}[p]
\resizebox{\hsize}{!}{\includegraphics{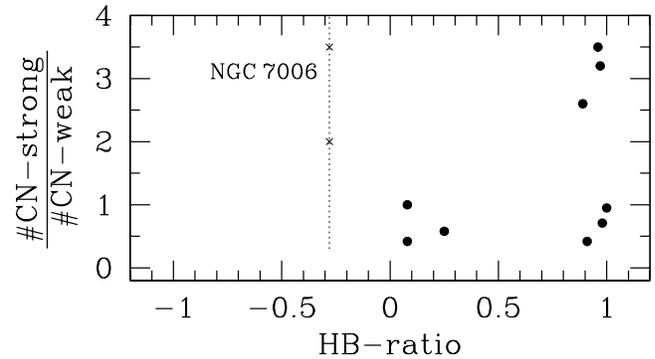}}
\caption{Comparison
        of the HB morphology and the content of CN-enhanced stars in
        GCs. We plot the HB morphology index
        $=\frac{\#B-\#R}{\#B+\#V+\#R}$ vs. the number ratio $r$
        between CN-enhanced and CN-weak stars. The crosses represent
        two different estimates of $r$ in \ngc (see text). The dotted
        line indicates the range $r$ in \ngc consistent with the
        uncertainties of the number counts. }
\label{fig_7006_hbcn}
\end{figure}

Fig.~\ref{fig_7006_hbcn} shows the HB index vs. the CN-ratio $r$ for
the sample of GCs with metallicities comparable to \ngc. Due to the
ambiguity of $r$ in \ngc we refrain from boldly plotting it in this
diagram. Instead, we indicate the range of $r$ that is consistent with
our data sample.  In this sample, no clear correlation between the HB
morphology and the ratio of CN-strong to CN-weak stars is apparent, as
also found in a similar study by \citet{smith02a}. M\,2 and NGC\,1904
fulfill the expectation of the rotation plus mixing scenario that GCs
with extremely blue HB morphology should have more CN-strong stars
than GCs with a redder HB morphology. But at the same time there are
GCs with very blue HB morphologies and low number ratios $r$
comparable to GCs with normal HB morphologies. \ngc seems to have too
large a ratio $r$ to support the scenario of an $r$-HB-type
correlation. If the distribution of angular momentum among stars in a
GC is correlated with the fraction of CN-enhanced stars, stellar
rotation cannot be identified as a second parameter of the HB
morphology, at least for our sample of GCs with metallicities of
[Fe/H]$\sim-1.6$~dex. We would like to point out that a correlation
between the distribution of CN absorption strengths and stellar
rotation in GCs has still to be proven. Since the initial distribution
of angular momentum can only be measured on the main-sequence of GCs
with high-resolution spectroscopy, limitations in telescope size and
instrumentation have not permitted such studies so far. While
main-sequence stars in GCs such as M\,3 and M\,13 are now in reach of
large telescopes, the distant cluster \ngc is still too far away for
efficient studies of such stars.

\subsection{An anti-correlation between CN and CH?}

\begin{figure}
 \resizebox{\hsize}{!}{\includegraphics{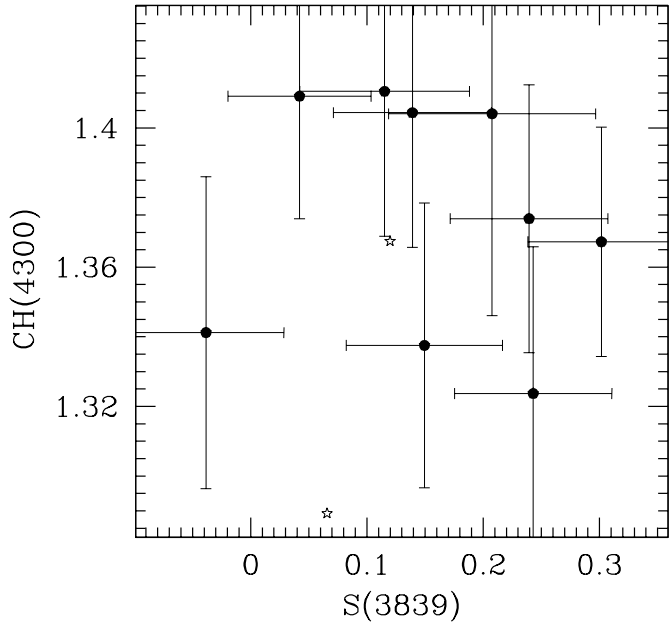}}

\caption{Comparison 
        of the CN and CH absorption strengths in \ngc. Stars with M$_V
        \ge -2.3$~mag are plotted with filled circles.  The two
        brightest giants in the sample are plotted with open stars.  } 
        \label{fig_7006_cnch}
\end{figure}

For a variety of GCs with CN abundance inhomogeneities an
anti-correlation between the CN band and CH G-band absorption
strengths has been detected, e.g., in M\,2, M\,3, M\,10, M\,13, and
NGC\,6752. Such a CN-CH anti-correlation is expected if CN variations
are produced by nucleosynthesis of material in the CNO cycle. Nitrogen
would be enriched at the cost of carbon and oxygen. The N-enhancement
will lead to CN-strong stars, while the reduced carbon content in such
stars would suppress CH molecule formation.  In
Fig.~\ref{fig_7006_cnch} we plot the \scnb vs. \sch spectral index for
stars in the narrow luminosity range  with $M_V\ge-2.3$~mag.  The
impact of temperature effects on the molecule formation is therefore
expected to be small. The uncertainties in the CH indices, coupled
with an apparently small range in CH(4300) values among our \ngc
giants, do not allow a quantification of the true scatter in CH
absorption strength.

Three of our stars (III-40, II-4, and III-33) are in common with the
sample investigated by \citet{friel82}, who determined carbon
abundances for these stars from low-resolution spectroscopy (included
in Tab.~\ref{tab_7006_indices}). We find a reasonable correlation
between our measured CH(4300) band strengths and the \citet{friel82}
carbon abundance: Star III-33 --- having the lowest [C/Fe] abundance
among the three stars --- also has the lowest measured CH absorption
band strength. On the other hand, star III-33 also has the lowest
S(3839) absorption strength, which also appears to be consistent with
Fig.~5a of \citet{friel82}.  We conclude that there is no evidence of
a CN-CH anti-correlation from Fig.~\ref{fig_7006_cnch}, although it
might be hidden in the observational uncertainties.

\subsection{Radial distribution of CN strengths}

\begin{figure}
\resizebox{\hsize}{!}{\includegraphics{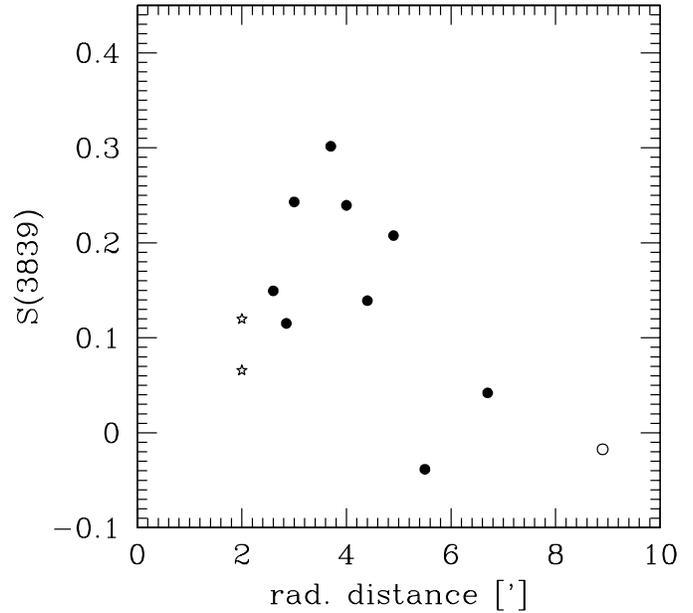}}

 \caption{A plot of the CN absorption strength of \ngc giants
   vs. projected distance to the cluster center.  Stars with strong CN
   bands are found at small radial distances, while the two CN-weak
   stars have large radial distances. Note that there are no CN-weak
   stars with a radial distance $\le 5'$. Only stars with $M_V\ge-2.3$
   are plotted with filled circles. The open circle indicates the
   location of the field star III-1.  The two brightest giants are
   plotted with open stars. \label{fig_7006_radplot}}
\end{figure}

\ngc has been reported to show --- besides its global second parameter
effect --- an internal ``second'' parameter effect. The morphology of
the HB changes with the radial distance to the cluster's center
\citep{buonanno91}: The HB morphology appears bluer in the central
parts than in the outskirts. It appears therefore interesting to see
if a similar trend can be found for the distribution of CN absorption
strength.  Our sample of eleven \ngc stars lacks the statistical
significance of a large survey. Nevertheless, we investigate the
radial distribution of CN-weak and CN-strong stars in our sample.  In
Fig.~\ref{fig_7006_radplot} we plot the CN index S(3839) vs. the
projected radial distance from the cluster center on the sky.
Surprisingly, all CN-enhanced stars have distances smaller than
$4.5$\arcmin, while the two CN-weakest stars have larger distances. A
radial distance of $4.5$\arcmin corresponds to approximately $0.7$
tidal radii ($r_t=6.34$\arcmin; \citealt{harris96}). The lack of
CN-strong stars at large radial distances is an uncertain result due
to the small number of observed stars; the absence of CN-poor stars in
the cluster's inner part is worth noting. We test the significance of
a difference in the radial distributions of CN-strong
(S(3839)$\ge0.1$) and CN-weak (S(3839)$\le0.1$) stars with a
Kolmogorov-Smirnov test. This test results in a $0.6$\% probability
that CN-strong and CN-weak stars follow the same radial distribution.
 It is worth mentioning here that \citet{buonanno91} found the
  radial gradient in HB morphologies on a scale of order of 70'',
  while we find a lack of CN-strong stars on radial scales of order
  4'. A detailed study of CN -- especially in the central parts --
  appears desirable.

A radial gradient in the distribution of CN strengths has been found
in the globular cluster 47\,Tuc \citep{norris79}. In this cluster,
CN-strong stars are more centrally concentrated than CN-weak stars.
\citet{briley97} pointed out that there are no CN-weak stars in the
very center of 47\,Tuc. The same situation might be seen in
\ngc. Although the gradient in \ngc has to be proven by a larger
sample, it would not be without precedent.

If this gradient in CN absorption strength is indeed real, its
implication is not unique. In the scenario of rotationally induced
mixing this gradient would reflect a gradient in the angular momentum
distribution of the stars, where stars in the inner part of
\ngc would have higher angular momentum. Accordingly, the central
stars would experience enhanced internal mixing and enhanced mass
loss.  It is not clear if the distribution of angular momentum dates
to the star formation epoch of \ngc, or if close stellar encounters
with transfer of angular momentum lead to a spin-up of those stars
orbiting preferentially within the inner parts of the GC.

In other scenarios, such as accretion of CNO processed stellar winds
(e.g., investigated in detail by \citealt{thoul02}), the gradient in
CN strengths would reflect different efficiencies of the accretion
process.  Stellar winds may settle at the center of a GC; if so, stars
with orbits restricted to the central regions would accrete more
material than stars on more extended orbits. Thus, the central
concentration of CN-enriched stars could be consistent with the
accretion scenario. A more detailed investigation of the suggested
gradient in CN absorption strengths in \ngc appears desirable.

\subsubsection{Sodium, Oxygen, and CN among NGC 7006 giants}

In GCs with CN abundance variations, a positive correlation between
the sodium abundance and CN is often found. At the same time, CN and
the oxygen abundance tend to be anti-correlated (e.g., the review by
\citealt{kraft94}). An oxygen-CN anti-correlation is a natural
consequence of CNO-processing. The situation is less clear in the case
of the sodium-CN correlation: initial interpretation of this
correlation was that material in CN-strong GC red giants had been
nucleosynthesized in more massive stars, whose stellar ejecta were
accreted by those stars seen at the present time. However, a neon
$\rightarrow$ sodium proton capture chain was identified as a possible
mechanism to allow internal sodium enhancement even in low-mass stars
\citep{dd90,langer93,cavallo96,cavallo98}. If this process is indeed 
present in RGB stars, the internal mixing scenario would still be
consistent with a sodium-CN correlation as has been argued by a number
of authors (e.g., \citealt{weiss00}).

\begin{figure*}

\parbox{7.5cm}
  {\includegraphics[width=7cm]{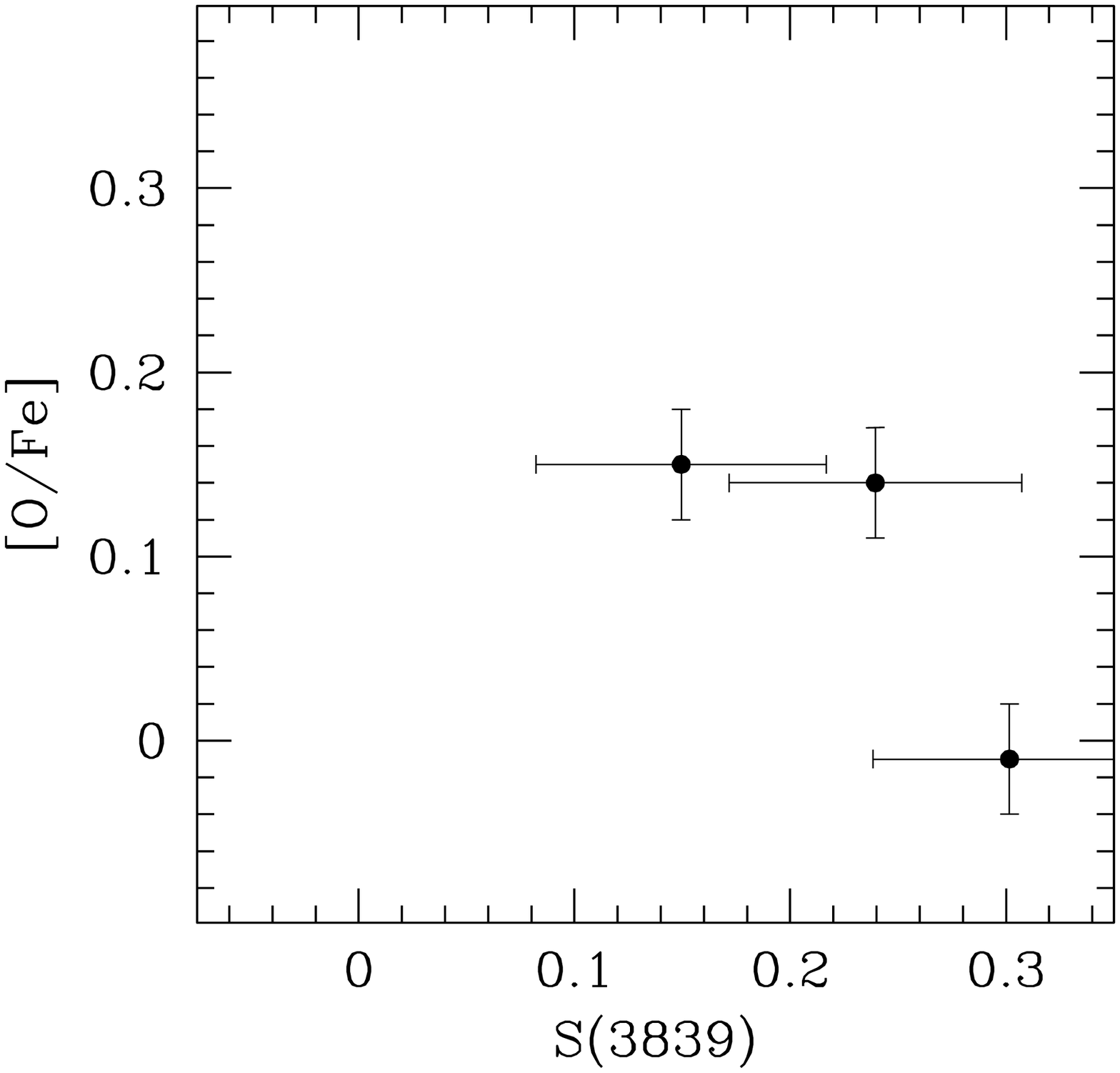}}
\hfill
\parbox{7.5cm}
 {\includegraphics[width=7cm]{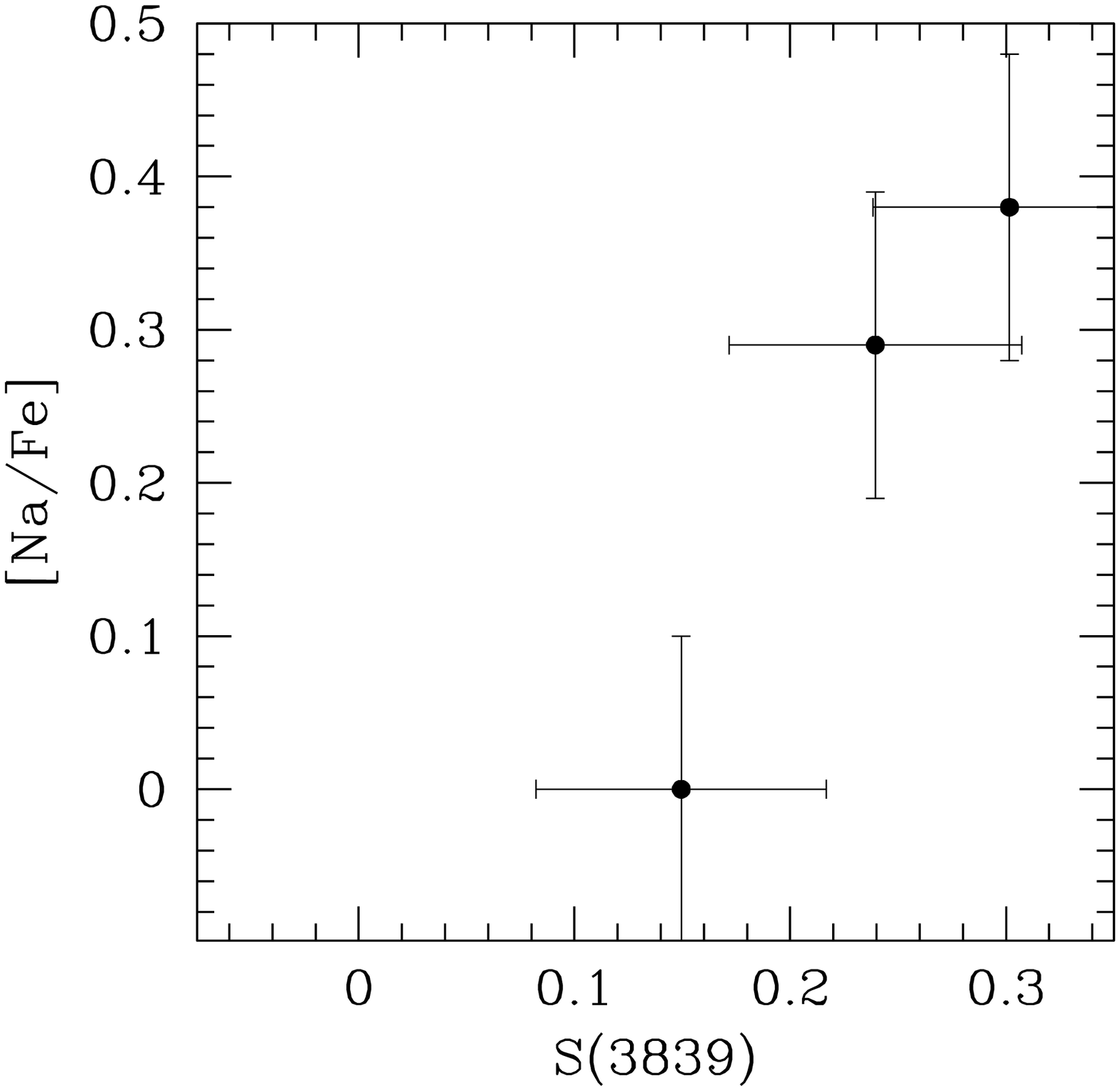}}

\caption{Comparison
  of CN S(3839) band strength with oxygen (left) and sodium (right)
  abundance measurements for stars in common with
  \citet{kraft98}. Within the strong limitation of the dataset, an
  oxygen-CN anti-correlation and a positive sodium-CN correlation is
  consistent with the observations. \label{fig_7006_sodium}}
\end{figure*}

Does \ngc obey the general correlations between CN, sodium, and oxygen
observed in other GCs?  In Fig.~\ref{fig_7006_sodium} we compare the
CN absorption strengths measured in this study to the sodium and
oxygen abundances measured by \citet{kraft98} from high-resolution
spectroscopy obtained at the Keck telescope. There are five stars in
common with this study, including the two peculiar variable stars
which are excluded from this comparison. The three remaining stars
with measured oxygen and sodium abundances are III-33, III-48, and
III-46. All stars are intermediate or strong in CN; no CN-poor stars
are included in this common sample. Although lacking strong
significance, a CN-oxygen anti-correlation as well as a CN-sodium
positive correlation is suggested.

\section{Summary}

Based on single-slit spectroscopy at the 3m Shane telescope obtained
in seven nights of observing time we have measured CN and CH molecule
absorption strengths for twelve stars in \ngc.  We discussed the
variations in CN and CH in the context of the second parameter effect
of horizontal branches in GCs. Within the limitations of this modest
set of CN observations, we found indications of (i) a continuous
distribution of CN S(3839) index values, (ii) a range in CN absorption
strength among RGB stars comparable to the CN spread present in the
GCs M\,3 and M\,10, and (iii) a larger fraction of RGB stars seem to
be CN-enriched in \ngc than in the two comparison clusters. A radial
gradient in the CN enrichment of stars in \ngc is suggested.

At a metallicity of [Fe/H]$\sim-1.6$~dex, we found that the range of
CN strengths does not differ among GCs of different HB morphologies,
as might have been expected in a scenario of internal nucleosynthesis
and subsequent dredge-up due to rotationally induced mixing. Thus,
stellar rotation could not be clearly identified as a second parameter
of horizontal branch morphology.  However, the small sample of stars
observed in \ngc does not allow us to completely disprove this scenario.

One caveat is necessary to state.  If the distribution of angular
momentum is a second parameter, it might not necessarily be traced in
a sufficient way by CN variations on the RGB, since an abundance
spread due to internal stellar evolution could be hidden by stronger,
primordial CN variations. In fact, recent detections of CN abundance
variations on the faint main-sequence of 47\,Tuc  strongly suggest that
 abundance variations due to external pollution effects might exceed the 
 imprint of internal stellar evolution \citep{harbeck03}. In so far it is not
 surprising that we can not link rotationally induced mixing to the problem
  of the HB morphology in this paper.

A direct measurement of the angular momentum distribution (i.e., the
distribution of $v\sin i$) in GCs at the main-sequence turn-off (MSTO)
is the more appropriate method to investigate the effect of rotation
on stellar evolution\footnote{As bright RGB stars have very extended
envelopes, conservation of angular momentum requires extremely low
surface rotation speeds of these stars; the broadening of absorption
lines in their spectra cannot be used to determine $v \sin i$.}. In
particular, a measurement of the angular momentum distribution would
allow a quantitative approach to the connection between rotation and
HB morphology. Fiber-fed spectrographs at large telescopes (i.e.,
FLAMES at the VLT) could allow measurement of rotation speeds of the
faint MS stars in nearby globular clusters.

\begin{acknowledgement}
GHS gratefully acknowledges the support of NSF grant AST 00-98453.
\end{acknowledgement}

\bibliographystyle{aa}

\clearpage

\clearpage

\begin{table}

\begin{tabular}[t]{lll} 

\hline \hline 
Star-ID & Date & Exposure Time \\ \hline

II-18  & 19/09/96 & $3 \times 2700$~s \\
II-46  & 19/09/96 & $3 \times 2700$~s \\
V-19   & 20/09/96 & $3 \times 2700$~s \\
V-54   & 20/09/96 & $1 \times 2700$~s \\
II-103 & 02/08/97 & $2 \times 3600$~s \\
II-89  & 02/08/97 & $2 \times 3600$~s + $1 \times 2700$~s \\
III-1  & 03/08/97 & $1 \times 3600$~s + $1 \times 1800$~s \\
II-85  & 03/08/97 & $2 \times 3600$~s + $1 \times 2700$~s \\
II-4   & 25/08/97 & $3 \times 3600$~s\\
III-46 & 25/08/97 & $2 \times 3600$~s + $1 \times 2400$~s\\
III-33 & 04/07/00 & $2 \times 3600$~s + $2 \times 2700$~s\\
III-40 & 05/07/00 & $3 \times 3600$~s + $1 \times 2700$~s\\ \hline
\end{tabular}

\caption{Logbook of Observations.}
\label{tab_7006_obslog}

\end{table}

\begin{table}

\begin{tabular}[t]{lccccc}
 \hline \hline 
 Star-ID &  V &  (B-V) &  \scnb &   \sch & [C/Fe]$^1$\\ \hline

 \input harbeck.tab02
 \hline
\end{tabular}
\caption{Parameters of program stars. $^1$adopted from \citet{friel82}}
\label{tab_7006_indices}
\end{table}

\begin{table}
\begin{tabular}[b]{lllll}

\hline \hline

 Cluster & [Fe/H] & HB & r & Sources \\ 

\hline

M\,10     & -1.52  &  0.98 & 0.71 & 1,2 \\
M\,13     & -1.54  &  0.97 & 3.2  & 3 \\
M\,3      & -1.57  &  0.08 & 0.42 & 4 \\
NGC\,3201 & -1.58  &  0.08 & 1.00 & 5 \\
NGC\,1904 & -1.57  &  0.89 & 2.60 & 6 \\
M\,22     & -1.64  &  0.91 & 0.42 & 7 \\
NGC\,6752 & -1.56  &  1.00 & 0.95 & 8 \\
NGC\,6934 & -1.54  &  0.25 & 0.58 & 9 \\
M\,2      & -1.62  &  0.96 & 3.5  & 10\\ 
NGC\,7006 & -1.63  & -0.28 & {\small (2-3.5)}  & this work \\ \hline

\end{tabular}

\caption{ [Fe/H] and 
   HB morphologies from \citet{harris96}; $r$ is the number ratio
   $r=\frac{\#CN-strong}{\#CN-weak}$. References: 1 --
   \citet{briley93}; 2 -- \citet{smith97}, 3 --- \citet{norris87}, 4
   -- \citet{smith02}, 5 -- \citet{smith82}; 6 -- \citet{langer92}, 7
   -- \citet{norris83}, 8 -- \citet{norris81}; 9 -- \citet{smith86},
   10 -- \citet{smith90}}
   \label{tab_7006_cn}

\end{table}

\end{document}

%% file: harbeck.tab02.tex
II-103 & 16.23 &  1.41 & 0.302 $\pm$ 0.06 & 1.367 $\pm$  0.03 \\ 
II-18 & 16.32  &  1.33 & 0.149 $\pm$ 0.07 & 1.338 $\pm$  0.04 \\ 
II-4 & 16.62   &  1.21 & 0.139 $\pm$ 0.07 & 1.404 $\pm$  0.04 & -0.82\\ 
II-46 & 16.38  &  1.38 & 0.239 $\pm$ 0.07 & 1.374 $\pm$  0.04\\ 
II-85 & 16.60  &  1.20 & 0.115 $\pm$ 0.07 & 1.410 $\pm$  0.04 \\ 
II-89 & 16.53  &  1.21 & 0.243 $\pm$ 0.07 & 1.324 $\pm$  0.04 \\ 
III-1 & 16.44  &  1.37& -0.017 $\pm$ 0.07 & 1.428 $\pm$  0.04 \\ 
III-33 & 16.77 &  1.10& -0.038 $\pm$ 0.07 & 1.341 $\pm$  0.05 & -0.97\\ 
III-40 & 16.96 &  1.12 & 0.208 $\pm$ 0.09 & 1.404 $\pm$  0.06 & -0.75 \\ 
III-46 & 16.78 &  1.24 & 0.042 $\pm$ 0.06 & 1.409 $\pm$  0.04 \\ 
V-19 & 15.50   &  1.72 & 0.120 $\pm$ 0.07 & 1.367 $\pm$  0.04\\ 
V-54 & 15.85   &  1.35 & 0.066 $\pm$ 0.05 & 1.289 $\pm$  0.03 \\ 